\documentclass[11pt,twoside
]{article}

\usepackage{baaa2010}
\usepackage{graphicx}
\usepackage{subfigure}
\usepackage{psfrag}
\usepackage{amssymb}
\usepackage[spanish,activeacute,english]{babel}
\usepackage[latin1]{inputenc}
\usepackage[T1]{fontenc} 
\usepackage{ae,aecompl} 
\usepackage{latexsym}
\usepackage{verbatim}
\usepackage{amsmath}
\usepackage{stmaryrd}
\usepackage{amsfonts}
\usepackage{amssymb}
\usepackage{wasysym}
\usepackage[colorlinks=true,dvips]{hyperref}

\begin{document}
\myselectenglish
\vskip 1.0cm
\markboth{ De Rossi, Tissera \& Pedrosa }%
{Supernova feedback and the Tully-Fisher relation}

\pagestyle{myheadings}
\vspace*{0.5cm}
\noindent PRESENTACIÓN ORAL
\vskip 0.3cm
\title{The role of supernova feedback on the origin of
the stellar and baryonic Tully-Fisher relations}


\author{M.E. De Rossi$^{1,2,3}$, 
P.B. Tissera$^{1,2}$, S.E. Pedrosa$^{1,2}$}

\affil{%
  (1) Instituto de Astronomía y Física del Espacio (IAFE)\\
  (2) CONICET \\
  (3) Facultad de Ciencias Exactas y Naturales - UBA\\
}

\begin{abstract} 
In this work, we studied the stellar and baryonic Tully-Fisher relations
by using hydrodynamical simulations in a cosmological framework.
We found that supernova feedback plays an important role on shaping the stellar Tully-Fisher
relation causing a steepening of its slope at the low-mass end, consistently with observations.
The bend of the relation occurs at a characteristic velocity of approximately $100 \ {\rm km} \ {\rm s}^{-1}$, in concordance with
previous observational and theoretical findings.
With respect to the baryonic Tully-Fisher relation, the model predicts a linear trend
at $z \sim 0$ with a weaker tendency for a bend at higher redshifts.  
In our simulations, this behaviour is a consequence of the more efficient action of supernova feedback at regulating the
star formation process in smaller galaxies.
\end{abstract}

\begin{resumen}
En este trabajo, estudiamos las relaciones de Tully-Fisher estelar
y bariónica mediante la utilización de simulaciones hidrodinámicas dentro de un marco cosmológico.
Encontramos que la retroalimentación al medio por supernovas juega un rol importante en la determinación
de la forma de la relación de Tully-Fisher estelar incrementando su pendiente en
el sector menos masivo, consistentemente con las observaciones.  El cambio de pendiente de la relación ocurre
para una velocidad característica de aproximadamente $100 \ {\rm km} \ {\rm s}^{-1}$, en concordancia con hallazgos observacionales y teóricos
previos.
Con respecto a la relación de Tully-Fisher bariónica,  el modelo 
predice un comportamineto lineal a $z \sim 0$ con una tendencia débil a curvarse hacia
corrimientos al rojo altos.
En nuestras simulaciones,  este comportamiento es una consecuencia de la acción más eficiente de la retroalimentación
al medio por supernovas
en la regulación de la formación estelar de las galaxias más pequeñas.
\end{resumen}

\section{Introduction}
The correlation between the luminosity and the rotation velocity for spiral
galaxies (TFR, Tully \& Fisher 1977) has been widely studied during the last decades.
However, it is now accepted that the TFR is a result of the more fundamental scaling  relation between
the rotational velocity of a galaxy and its
stellar mass (sTFR) or
baryonic mass (bTFR).

In particular, there is evidence that the sTFR exhibits a change in its slope at around
 $\sim 90 \ {\rm km} \ {\rm s}^{-1}$ 
in the sense that slow-rotators have lower stellar masses than
those derived from the extrapolation of the linear fit for fast-rotators (McGaugh et al. 2000;
Amorín et al. 2009). Moreover, McGaugh et al. 2010 have recently reported that
there is also a bend in the bTFR but located at a lower rotation velocity.

In this work, we performed hydrodynamical simulation in a $\Lambda$CDM universe to study the
role of supernova (SN) feedback on the origin of the shape of the sTFR and bTFR and on their evolution
with redshift.

\section{Simulations and sample selection}
We performed numerical simulations consistent with the concordance $\Lambda$CDM
universe with ${\Omega}_{\rm m}=0.3$, ${\Omega}_{\Lambda}=0.7$, ${\Omega}_{\rm b}=0.04$
and ${\rm H}_{0} = 100  \, h^{-1} \, {\rm km} \, {\rm s}^{-1} \, {\rm Mpc}^{-1}$ with $h=0.7$.
A version of the chemical code GADGET-3 including treatments for metal-dependent
radiative cooling, stochastic star formation, a multiphase model for the ISM and SN feedback
(Scannapieco et al. 2006) was employed.
The simulated volume corresponds to a cubic box of a comoving
10 Mpc $h^{-1}$ side length.  The simulation has a mass resolution of
$5.93 \times 10^6 M_{\odot} h^{-1}$ and $9.12 \times 10^5 M_{\odot} h^{-1}$ for the dark and
gas phase, respectively.

Simulated disc-like galaxies were identified following the methods describe by De Rossi et al. (2010).
The mean properties of galactic systems were estimated
at the baryonic radius ($R_{\rm bar}$), defined as the one which encloses 83 per cent of the
baryonic mass of the systems.
We found that the tangential velocity of these systems constitutes a good representation
of their potential well so that, for the sake of simplicity, we used the circular velocity
estimated at $R_{\rm bar}$ as the kinematical indicator for our study.

\section{Results and discussion}

With respect to the local sTFR, fast-rotators ($100 \ {\rm km} \ {\rm s}^{-1} < V < 250 \ {\rm km} \ {\rm s}^{-1}$)
describe a linear relation of the form
$\log (M_{*} / M_{\odot} h^{-1}) =$ $(3.68 \pm 0.09)$ $\log (V / \ 100 \ {\rm km} \ {\rm s}^{-1})$ $+ (9.42 \pm 0.26)$,
which is in general good agreement with observations  (e.g. McGaugh et al. 2000).
Moreover, our model is also able to reproduce the observed steepening of the relation in the case of
slow-rotators ($V < 100 \ {\rm km} \ {\rm s}^{-1}$; see De Rossi et al.  2010 for details), in agreement
with previous observational and theoretical works (Larson 1974; Dekel \& Silk 1986; McGaugh et al. 2000).
In order to analyse at what extent SN feedback might be responsible for the bend
of the sTFR, we suppress this mechanism from our simulations.  We found that, when SN feedback is
turned-off, a single linear relation is recovered  with a flatter slope ($\approx 2.97$), suggesting the important role
of SNs in the determination of the shape of the sTFR.

In the case of the local bTFR, we obtained a linear trend for the whole range of circular
velocities resolved by these simulations ($40 \ {\rm km} \ {\rm s}^{-1} < V < 250 \ {\rm km} \ {\rm s}^{-1}$).
This result does not disagree with the findings of McGaugh et al. (2010) since they 
reported that the bend of the bTFR is at a lower rotation velocity ($V \sim 20 \ {\rm km} \ {\rm s}^{-1}$).
Furthermore, the simulated bTFR follows a relation of the form 
$\log (M_{b} / M_{\odot} h^{-1}) = (3.23 \pm 0.08) \log (V / \ 100 \ {\rm km} \ {\rm s}^{-1}) + (9.56 \pm 0.23)$,
consistently with observations for late and early type galaxies (De Rijcke et al. 2007; Gurovich et al. 2010).

By analysing the simulated sTFR and bTFR within the redshift range $0<z<3$, 
we found that neither of them show a significant evolution
of the slope at the high-mass end.  However, in the same redshift range, 
we detected an increase of the zeropoint by $\sim 0.44$ dex and $\sim 0.33$ dex for the sTFR and bTFR, respectively.

In order to explore how these trends arise in our SN model which does not
introduce scale-dependent parameters,  we analysed its effects
on the gas-phase of simulated galaxies.
Firstly, we investigated the importance of SN-driven outflows by calculating
the fraction $f_{i} = {\Omega}_{\rm m} M_{i}  / {\Omega}_{b} M_{\rm vir}$
for each  galaxy at $z=0$, where $M_{\rm vir}$ is the total mass within the virial radius
and $M_{i}$ denotes the stellar ($f_{*}$) or the baryonic ($f_{b}$) component within $R_{\rm bar}$.
Our results indicate that $f_{*}$ is an increasing function of the circular velocity suggesting that SN feedback might
be strongly regulating the star formation efficiency (eSFR) of smaller galaxies.  To analyse this issue, we
compared the correlation between the eSFR and the circular velocity
of the systems in the case of the SN-feedback model with the one obtained from  a feedback-free run.  As expected, the
SN feedback model predicts lower eSFRs at a given velocity with
the stronger differences at the low-mass end of the relation, consistently with
the steepening of the sTFR for slow-rotators.
By studying the fraction $f_{b}$ as a function of the circular velocity, we found a weaker
correlation than in the case of $f_{*}$,  which accounts for the trend to recover a single slope for the bTFR.  However, the
fact that $f_{b} < 0.6$ for the whole sample suggests that galactic winds are important over the 
entire range of simulated stellar masses.
Finally, we estimated the fraction $f_{b}^{\rm vir}$ defined as the ratio between the 
simulated baryonic mass within the virial radius and the expected one. By comparing
$f_{b}$ and $f_{b}^{\rm vir}$, we determined  that an important 
amount of the missing baryons within galaxies can be found in the surrounding halo (a detailed discussion can be found in De Rossi et al. 2010).  
However, given that $f_{b}^{\rm vir}$ is within the range $0.1 - 1$ for the whole sample, 
it is clear that a significant fraction of the gas-phase of simulated galaxies is blown away as a consequence of very 
efficient galactic winds. And, these effects are more prominent in smaller systems.

Basically, in our model, star formation is triggered mainly when the gas gets cold and dense
\footnote{In this simulations, the cold phase is defined as the gas component with temperature $T < T_c $ where $ T_c = 8\times 10^4$ K
and density $\rho > 0.1 \rho_{\rm c}$ where $\rho_{\rm c} $ is $7 \times 10^{-26} {\rm g \ cm^{-3}}$. 
Otherwise, the gas is classified as hot phase.
The reader is referred to
Scannapieco et al. 2006 for more details about the model.}.
As a consequence of star formation, SN energy is released heating up the surrounding cold ISM.  
The smaller virial temperatures of slow rotators lead to a more efficient transition of the gas from 
the cold to the hot phase, and hence, generates a decrease of the star formation activity in these systems.  
However, the cooling times for these galaxies are still too short compared to the dynamical times to allow this hot phase to 
be stable and the gas can return to the cold phase in short time-scales.  
In fact, SN feedback leads to a self-regulated cycle of heating and cooling generating an important 
regulation of the star formation process in smaller galaxies.  
In the case of fast rotators,  the hot phase is established at higher temperature due to the continuous injection
of SN energy and the increase of the 
virial temperature of the dark halos hosting them.  And, since in these systems the cooling times get longer compared 
to the dynamical times, the hot gas is able to remain in this phase.  
However, at this stage, galactic winds will be more difficult to be triggered since the SN energy 
accumulated by the cold phase is not enough to match that of its nearby hot environment, 
which is a condition required by the model to promote particles to the hot phase.
Meanwhile, this gas remains available for star formation and, consequently,
SN feedback is not efficient at regulating the star formation in larger galaxies.
Interestingly, this model predicts that the transition from efficient to inefficient gas cooling occurs at 
around the same characteristic velocity of $100 \ {\rm km} \ {\rm s}^{-1}$ where the sTFR bends 
and is also in agreement with previous observational and theoretical findings.

\section{Conclusions}
We have studied the sTFR and bTFR by using cosmological simulations in a $\Lambda$CDM universe.
Our results suggest that SN feedback seems to be crucial to reproduced the observed bend of the sTFR 
as a consequence of the more efficient action of SN feedback in the regulation of the star formation activity 
in smaller galaxies.
Without introducing scale-dependent parameters, the model predicts that the bend occurs at a characteristic 
velocity of $\sim 100 \ {\rm km} \ {\rm s}^{-1}$, consistently with previous observational and theoretical works.
The reader is referred to De Rossi et al. (2010) for more details about this work.

\acknowledgements
MEDR 
thanks the Argentinian Astronomical Society for its partial financial support to attend this meeting.
We acknowledge support from the  PICT 32342 (2005) and
PICT 245-Max Planck (2006) of ANCyT (Argentina).
Simulations were run in Fenix and HOPE clusters at IAFE and CeCAR cluster at  University
of Buenos Aires.

\begin{referencias}

\reference Amor\'{\i}n, R., Aguerri, J.~A.~L., Mu\~noz-Tu\~n\'on, C. \&  Cair\'os L.~M. 2009, A\&A, 501, 75

\reference Dekel, A. \& Silk, J. 1986, ApJ, 303, 39

\reference De Rijcke, S., Zeilinger, W.~W., Hau, G.~K.~T., Prugniel, P. \& Dejonghe, H.
2007, ApJ, 659, 1172

\reference De Rossi, M. E., Tissera, P. B., Pedrosa, S. E. 2010, A\&A, 519, 89

\reference 
Gurovich, S., Freeman, K. C., Jerjen, H., Staveley-Smith, L., \& Puerani, I. 2010, AJ,  140, 663

\reference Larson, R. B. 1974, MNRAS, 169, 229

\reference McGaugh, S.~S., Schombert, J.~M., Bothun,
G.~D., \& de Blok, W.~J.~G. 2000, ApJ, 533, L99

\reference McGaugh, S.~S., Schombert, J.~M., de Blok,
W.~J.~G., \& Zagursky, M.~J. 2010, ApJ, 708L, 14

\reference Scannapieco, C., Tissera, P.B., White, S.D.M. \& Springel, V. 2006,
MNRAS, 371, 1125

\reference Tully, R.~B., \& Fisher, J.~R. 1977, A\&A, 54, 661

\end{referencias}

\end{document}